\documentclass[11pt]{article}
\usepackage{fancyhdr}
\usepackage{amssymb}
\usepackage{url}
\usepackage{xspace,pifont,epsfig,euler,textcomp,rotating,wrapfig}
\usepackage{marvosym}
\usepackage[usenames]{color}
\newcommand{\EA}{{\em et al.}\xspace}
\newcommand{\IE}{{\em i.e.}\xspace}
\newcommand{\EG}{{\em e.g.}\xspace}
\newcommand{\FI}[1]{Fig.~\ref{#1}\xspace}
\newtheorem{thm}{Theorem}

\newtheorem{Lemma}[thm]{Lemma}
\newtheorem{Proposition}[thm]{Proposition}

\newcommand{\TB}{\vspace{-0.1ex}}\newcommand{\TiE}{\setlength{\itemsep}{-1ex}}
\newenvironment{TightItem}{\TB\TB\begin{itemize}\TiE}{\end{itemize}\TB}
\newenvironment{TightEnum}{\TB\TB\begin{enumerate}\TiE}{\end{enumerate}\TB}
\newenvironment{TightDesc}{\TB\TB\begin{description}\TiE}{\end{description}\TB}

\newcommand{\comment}[1]{}

\newcommand{\eps}{\varepsilon}
\newcommand{\mod}{\mbox{ \bf mod }}

\newcommand{\es}{\mbox{\Pisymbol{psy}{198}}}

\newcommand{\Oli}[1]{{#1}}
\newcommand{\Mat}[1]{#1}

\newcommand{\Eu}[1]{$#1$\raisebox{2pt}{\scriptsize\EUR}}
\newcommand{\Rbr}[1]{#1}
\newcommand{\sci}[2]{{\sc #1}[$ #2$]}
\newcommand{\dfs}{{\sc Dfs}\xspace}
\newcommand{\Dfs}[1]{{\sc Dfs}($ #1$)}
\newcommand{\iif}{{\bf if}\xspace}
\newcommand{\eelse}{{\bf else}\xspace}
\newcommand{\ffor}{{\bf for}\xspace}
\newcommand{\bec}{$\leftarrow$\xspace}
\newcommand{\arr}{\rightarrow}
\newcommand{\MiED}{{\sc Min-ED}\xspace}
\newcommand{\MaED}{{\sc Max-ED}\xspace}
\newcommand{\MiTR}{{\sc Min-TR}$_p$\xspace}
\newcommand{\MiTRo}{{\sc Min-TR}$_1$\xspace}
\newcommand{\MiTRt}{{\sc Min-TR}$_2$\xspace}
\newcommand{\MaTR}{{\sc Max-TR}$_p$\xspace}
\newcommand{\MaTRo}{{\sc Max-TR}$_1$\xspace}
\newcommand{\regsat}{{\sc 2Reg-Max-SAT}\xspace}
\newcommand{\msat}{{\sc Max-SAT}\xspace}

\title{\vspace*{3ex}\bf Approximating Transitivity in Directed Networks\\[6ex]}

\author{
Piotr Berman\thanks{Research partially done while visiting Dept. of Computer
Science, University of Bonn and supported by DFG grant Bo 56/174-1.}\\
Department of Computer Science \& Engineering \\
Pennsylvania State University \\
University Park, PA 16802 \\
Email: {\tt berman@cse.psu.edu} \\
\and
Bhaskar DasGupta\thanks{Supported by NSF grants DBI-0543365, IIS-0612044 and IIS-0346973.} \\
Department of Computer Science \\
University of Illinois at Chicago \\
Chicago, IL 60607-7053 \\
Email: {\tt dasgupta@cs.uic.edu} \\
\and
Marek Karpinski\thanks{Supported in part by DFG grants, Procope grant 31022, and
the Hausdorff Center grant EXC59-1.} \\
Department of Computer Science \\
University of Bonn \\
53117 Bonn, Germany \\
Email: {\tt marek@cs.uni-bonn.de}
}

\date{}

\begin{document}

\maketitle

\begin{abstract}
We study the problem of computing a minimum equivalent digraph
(also known as the problem of computing a strong transitive reduction) 
and its maximum objective function variant,
with two types of extensions.  First,
we allow to declare a set $D\subset E$ and require that a valid solution $A$
satisfies $D\subset A$ (it is sometimes called transitive reduction problem).
In the second extension (called $p$-ary transitive reduction),
we have integer edge labeling and we view two paths as equivalent if they have
the same beginning, ending and the sum of labels modulo $p$.  A solution
$A\subseteq E$ is valid if it gives an equivalent path for every original path.
For all problems we establish the following: polynomial time minimization of
$|A|$ within ratio 1.5, maximization of $|E-A|$ within ratio 2, MAX-SNP
hardness even of the length of simple cycles is limited to 5.
%%%%
% added -- Bhaskar
%%%%
Furthermore, we believe that 
the combinatorial technique behind the approximation algorithm for
the minimization version might be of interest to other graph connectivity
problems as well.
\end{abstract}
\vspace{10ex}
\thispagestyle{empty}
\pagebreak
\setcounter{page}{1}

\section{Introduction}

\subsection{Definitions and motivation}
\label{defimod}
Minimum equivalent digraph
is a classic computational problem
(cf. \cite{MT69})
with several recent extensions
motivated by applications in social
sciences, systems biology etc.

The statement of the equivalent digraph problem is simple.  For a digraph
$(V,E)$ the {\em transitive closure} of $E$ is relation
% $u\stackrel{E\ }{\arr}v$ $\equiv$
''$E$ contains a path from $u$ to $v$''.  In turn, $A$ is an
equivalent digraph for $E$ if (a) $A\subseteq E$, (b) transitive closures of $A$ and
$E$ are the same.

The assumption that the valid solutions are the equivalent digraphs of $E$
yields two different optimization
problems when we define two objective functions:
\MiED, in which we minimize $|A|$, and \MaED, in which we
maximize $|E-A|$. where $A$ is an equivalent digraph for $E$.

Skipping condition (a) yields {\em transitive reduction} problem which is
optimally solved by Aho \EA \cite{AGU72}.
% These names are a bit confusing
% because one would expect a {\em reduction} to be a subset and an {\em equivalent
% set} to be unrestricted, but transitive reduction was first discussed when the
% name {\em minimum equivalent digraph} was already introduced \cite{MT69}.
This could motivate renaming the equivalent digraph as a
{\em strong transitive reduction} \cite{P94}.

In the study of biological systems networks of interactions are considered,
\EG nodes can be genes and an edge $(u,v)$ means that gene $u$ {\em regulates}
gene $v$.  Without going into biological details, regulations may mean at
least two different things: when $u$ is {\em expressed}, \IE molecules of
the protein coded by $u$ are created, the expression of $v$ can be
{\em repressed} or {\em promoted}.  A path in this network is an indirect
interaction, and promoting a repressor represses, while repressing a
repressor promotes (biologists also used the term {\em de-repression}).
Interactions of such two types can appear in other contexts
as well, including social networks.  This motivates an extension of the
notion of digraph and its transitive closure
described in points \ding{172}-\ding{174} below.

Moreover, for certain interactions we have direct evidence, so an instance
description includes set $D\subset E$ of edges which have to be present
in every valid solution.  Formally, we define $A$ to be a valid solution
to instance $(V,E,\ell,D)$ as follows:

\vspace{-1.5ex}
\begin{dingautolist}{172}
\item
$\ell:~E\arr\Bbb{Z}_p$;
\vspace{-1.5ex}
\item
a path $P=(u_0,u_1,\ldots,u_k)$ has characteristic
$\ell(P)=\sum_{i=1}^k\ell(u_{i-1},u_i)$ {\bf mod} $p$;
\vspace{-1.5ex}
\item
$Closure_\ell(E)=\{(u,v,q):~\exists P$
in $E$
from $u$ to $v$ and
$\ell(P)=q\}$;
\vspace{-1.5ex}
\item
$A$ is a $p$-ary transitive reduction of $E$ with a required subset $D$ if
$D\subseteq A\subseteq E$ and $Closure_\ell(A)=Closure_\ell(E)$.
\end{dingautolist}
\vspace{-1.5ex}

Our two objective functions define optimization problems \MiTR
and \MaTR.
\vspace{-1.5ex}

\subsection{Earlier results}

The initial work on
\MiED
% the minimum equivalent digraph
by Moyles and Thomson
\cite{MT69} described
an efficient reduction to the case of strongly connected graphs and
an exact exponential time algorithm for the latter.

Several approximation algorithms for \MiED
were described, by Khuller \EA \cite{KRY94},
with approximation ratio $1.617+\eps$ and by Vetta \cite{V01}  with
approximation ratio 1.5.  The latter result did not have a full peer review,
however.

If edges have costs, we can minimize $c(A)$ within factor 2 using
an algorithm for minimum cost rooted arborescence \cite{E71,K72} of
Edmonds (who described it) and Karp (who simplified it).  We find minimum
cost in- and out- arborescence in respect to an arbitrary root $r\in V$.

Albert \EA \cite{ADDS} showed how to convert an algorithm for \MiED
with approximation ratio $r$ to an algorithm for \MiTRo
with approximation ratio $3-2/r$.  They have also shown a
$2+o(1)$-approximation for \MiTR
%%%%
% added -- Bhaskar
%%%%
{\em when $p$ is a prime}.  
%%%%
% added ends -- Bhaskar
%%%%
Other heuristics for these problems
were investigated in \cite{ADDKSZW07, KZSAD07}.

On the hardness side, Papadimitriou \cite{P94} formulated an exercise to show
that strong transitive reduction is NP-hard, Khuller \EA have proven it
formally and they also showed MAX-SNP hardness.  Motivated by their {\em cycle
contraction} method in \cite{KRY94}, they were interested in the complexity of
the problem when there is an upper bound $\gamma$ on the cycle length;
in \cite{KRY95} they showed that \MiED is polynomial with $\gamma=3$, NP-hard
with $\gamma=5$ and MAX-SNP-hard with $\gamma=17$.

\subsection{Results in this paper}

We show an approximation algorithm for \MiED with ratio 1.5,  We use a
method somewhat similar to that of Vetta \cite{V01}, but our combinatorial
lower bound makes a more explicit use of the primal-dual formulation
of Edmonds and Karp, and this makes it much easier to justify edge selections
within the promised approximation ratio.\footnote{
It appears that the approach of \cite{V01} may be correct, but
the proofs and the description of the algorithm seem to have
some gaps.
It is somewhat difficult to point out these gaps
without going into technical details; we will point out
some problems as an illustration in the proof section of the paper.}.

Next, we show how to modify that algorithm to approximate \MiTRo
within ratio 1.5.  One surely cannot use a method for \MiED as a ``black box''
because we need to control which edges we keep and which we delete.

We show approximation algorithm with ratio 2 for \MaTRo.
While it was shown
by Albert \EA~\cite{ADDKSZW07} that a simple greedy algorithm, delete
an unnecessary edge as long as one exists, yields ratio 3 approximation,
it is easy to provide an example of \MaED instance with $n$ nodes and
$2n-2$ edges in which greedy removes only one edge, and the optimum
solution removes $n-2$ edges.  Other known algorithms for \MiED are
not much better in the worst case when applied to \MaED.

We show that {\em for prime $p$} we can transform an equivalent digraph that contains
the required edges into a $p$-ary transitive reduction by a single edge
insertion per strongly connected component.   Because every $p$-ary transitive
reduction is also an equivalent digraph, this transformation implies
approximation
algorithms for \MiTR with ratio $1.5$ and for \MaTR with ratio $2$ (we can
compensate for the insertion of a single edge, so the ratio does not
change)
%%%%
% added -- Bhaskar
%%%%
\footnote{Albert \EA~\cite{ADDS} did
not use this approach and tried to approximate 
\MiTR for $p>1$ directly thus obtaining a 
$2+o(1)$-approximation for prime $p$.}.
%%%%
% added ends -- Bhaskar
%%%%

We simplify the MAX-SNP hardness proof for \MiED (the proof applies to \MaED
as well)
so it applies 
 even when $\gamma$, the maximum cycle length,
is 5.  This leaves open only the case of $\gamma=4$.
%
%Lastly, we show that the dynamic programming solution for solving exactly
%the Hamiltonian Cycle problem can be adapted, without adding much extra
%running time, to \MiED (this algorithm allows arbitrary edge costs).

\subsection{Some Motivations and Applications}

\paragraph{Application of \MiED: Connectivity Requirements in
Computer Networks.}
Khuller \EA \cite{KRY95} mentioned applications of \MiED to design of computer
networks that satisfy given connectivity requirements.

If a set of connections exists already, then this application motivates \MiTRo
(cf. \cite{KRZ99}).

\paragraph{Application of \MiTRo: Social Network Analysis and Visualization.}
\MiTRo can be applied to social network analysis and visualization. For example,
Dubois and C\'{e}cile~\cite{DB05} applies \MiTRo to the 
social network built upon interaction data (email boxes) of Enron corporation
to study
general properties (such as scale-freeness) of such networks and to help in
the visualization process. They use a
straightforward greedy approach which, as we have discussed, has inferior
performance, both for \MiTRo and \MaTRo.

\paragraph{Application of \MiTRt: Inferring Biological Signal Transduction
Networks.}
In subsection \ref{defimod} we motivated \MiTRt with the study of gene
regulatory networks.  The same issues apply to other
cellular interactions, like signal transduction networks and they
were addressed by
Albert \EA~\cite{ADDKSZW07}, with the use of an approximation algorithm 
\MiTRt.
% Most biological characteristics of a cell arise from complex interactions
% between its numerous constituents such as DNA, RNA, proteins and small
% molecules.  Cells use signaling pathways and regulatory mechanisms to coordinate
% multiple functions, allowing them to respond to and adapt to an
% ever-changing environment. Genome-wide experimental methods now identify
% interactions among thousands of cellular components.  Experimental information
% about the involvement of a specific component in a given signal transduction
% network can be partitioned into three categories.  First, biochemical evidence,
% that provides information on enzymatic activity or protein-protein interactions. 
% To synthesize all this information into a consistent network, we need to
% determine how the different pathways suggested by experiments fit together. 
% The \MiTRt problem is useful for determining the sparsest graph consistent with
% a set of experimental observations and it is the key
% part of the network synthesis procedure described in
% Figure~1 of Albert \EA~\cite{ADDKSZW07}.  

\subsection{Our Techniques} 

\noindent {\bf Approximation algorithm for the {\sc Min} objective.} 
Vetta used a primal/dual LP formulation for \MiED, more precisely,
a solution that satisfies a subset of linear constraints, and the optimum
solution for that subset is integer and it can be found using a maximum
matching.  We observed that a larger set of constraints also has this
property, and that the extra edges for justified by this extension 
make it much easier to analyze the algorithm.

To tackle \MiTRo problem we had to justify yet more edges, as the algorithm
is not allowed to delete the required edges.  We showed that we can use
a yet larger set of constraints, with a ``good enough'' solution that can
be found using a maximum weight matching.

We also used depth first search in a manner inspired by Tarjan's algorithm
for finding strongly connected components.
% consider a natural primal/dual IP formulation for \MiTRo that is a
% straightforward extension of a primal/dual formulation
% of minimum cost rooted arborescence problem on directed graphs
% (e.g., see~\cite{E71,K72} or~\cite[Corollary $6.12$]{KV00}). 
% This formulation introduces an inequality for every node set, but they
% exist for analysis only.
% We keep only some of the constraints and this gives a system whith
% a feasible integral solution $L$ of the dual which is not necessarily
% a feasible integral solution of the primal but nonetheless provides
% a lower bound on the optimal cost and can be computed in polynomial time. 
% This $L$ is subsequently converted into a correct (primal) solution
% using a careful DFS and amortized accounting---we are allowed to use 3 edges
% for every 2 edges of $L$---to satisfy the remaining constraints.
% This yields an $1.5$-approximation for \MiTRo that improved the
% $1.78$-approximation in~\cite{ADDS} and an
% alternate $1.5$-approximation for \MiED.

We also show an inherent limitation of our approach by showing an integrality
gap of the LP relaxation of the above IP to be of at least $4\over 3$. 

\noindent {\bf Approximation algorithm for the {\sc Max} objective.} 
For the \MaTRo, we utilize the integrality of the polytope of the rooted
arborescence problem to provide a $2$-approximation.  We also observe that the
integrality gap of the LP relaxation of the IP formulation is at least $3/2$. 

\noindent {\bf The $p$-ary case for prime $p$.}
We show that we can solve an instance of \MiTR/\MaTR by solving a
related instance of \MiTRo/\MaTRo and inserting a appropriately chooses single
edge.  In conjunction with our above results, this leads to
a $1.5$-approximation for \MiTR and $2$-approximation for \MaTR.
This method works only if $p$ is prime.

\noindent {\bf Inapproximability.}
We adapt the reduction used by Khuller \EA \cite{KRY94}, but we apply it to
a very restricted (and yet, MAX-SNP hard) version of \msat (cf. \cite{BKS03}).

\vspace{-1.5ex}
\subsection{Notation} 

We use the following additional notations.  
\begin{itemize}
\vspace{-1.5ex}
\item
$G=(V,E)$ is the input digraph;

\vspace{-1.5ex}
\item 
$\iota(U)=\{(u,v)\in E:~u\not\in U~\&~v\in U\}$,
$\iota(u_1,\ldots,u_k)= \iota(\{u_1,\ldots,u_k\})$;
; 

\vspace{-1.5ex}
\item 
$o(U)=\{(u,v)\in E:~u\in U~\&~v\not\in U\}$,
$o(u_1,\ldots,u_k)= o(\{u_1,\ldots,u_k\})$;

\vspace{-1.5ex}
\item 
$scc_A(u)$ is the strongly connected component containing vertex
 $u$ in the digraph $(V,A)$; 

\vspace{-1.5ex}
\item 
$T[u]$ is a the node set of the subtree with root $u$ (of a rooted tree $T$). 
\end{itemize}
\vspace{-1.5ex}

\section{A Primal-Dual Linear Programming Relaxation of TR$_1$}
\label{p1d1}

Moyles and Thompson \cite{MT69} showed that \MiED can be reduced in linear
time to the case when the input graph $(V,E)$ is strongly connected, therefore
we will assume that $(V,E)$ is already strongly connected.
In Section~\ref{general} we will show the same  for \MiTR.

The minimum cost rooted arborescence problem on $G$ is defined 
as follows. We are given a weighted digraph $(V,E)$, a cost function
$c:E\arr {\mathbb{R}}_+$
and root node $r\in V$. A valid solution is $A\subseteq E$ such that in $(V,A)$ there is
a path from $r$ to every other node and we need to minimize $c(A)$.
An LP formulation for this was provided by
Edmonds and Karp and goes as follows.
As in any other edge/arc selection problems, we use the linear space with a coordinate
for each arc, so edge sets can be identified with $0$-$1$ vectors, and for
an arc $e$ variable $x_e$ describes whether we select $e$ ($x_e=1$) or not
($x_e=0$). Then, the LP formulation is: 
\begin{tabbing}
01234,\=01234567890123456789012345678901234567890\= \kill
(primal P1) \\[0.5ex]
 minimize $c\cdot x$ subject to \\[0.5ex]
\> $x\ge 0$ \\[0.5ex]
$\,\iota(U)\cdot x \ge 1$ for all $U$ s.t. $\es\subset U\subset V$ and
$r\not\in U$ \>\> (2.1)
\end{tabbing}
Edmonds~\cite{E71} and Karp~\cite{K72} showed that the above LP always has an integral
optimal solution and that we can find it in polynomial-time.

We can modify the above LP formulation to a LP formulation for \MiED
provided we set $c(e)\equiv 1$ and in (2.1)
we remove ``and $r\not\in U$'' from the condition.
The dual program of this LP can be constructed by 
having a vector $y$ that has a coordinate $y_U$ for 
every $\es\subset U\subset V$; both the primal and the dual is written
down below for clarity: 
\begin{tabbing}
01234,\=
0123456789012345678901234567890123456\=
01234567,\=
\kill
(primal P2)\>\>(dual D2) \\[0.5ex]
minimize ${\bf 1}\cdot x$ subject to 
\>\> maximize ${\bf 1}\cdot y$ subject to \\[0.5ex]
\> $x\ge 0$
\>\> $y\ge 0$ \\[0.5ex]
$\,\iota(U)\cdot x \ge 1$ for all $U$ s.t. $\es\subset U\subset V$ 
\>\>
$\sum_{e\in\iota(U)} y_U\le 1$  for every $e\in E$
\end{tabbing}
We can change P2 into the
LP formulation for \MaED by replacing the objective to
``maximize ${\bf 1}\cdot({\bf 1}-x)$''.
and the dual is changed accordingly to reflect this change. 

From now on, by a {\em requirement} we mean a set of edges $R$
such that a valid solution
must intersect it; in LP formulation it means that we have a constraint
$Rx\ge 1$.

We can extend P2 to an LP formulation for TR$_1$ 
by adding a one-edge requirement $\{e\}$ (\IE inequality $x_e\ge 1$) for each
required edge $e$.

We can obtain a lower bound for solutions of P2 by solving P3, an IP obtained
from P2 by allowing only those
requirements $Rx\ge 1$ that for some node $u$ satisfy
$R\subseteq\iota(u)$ or $R\subseteq o(u)$.  To find requirements of P3
efficiently, we
first find strongly connected components of $V-\{u\}$. Then, 
\begin{TightDesc}
\item[(a)] for each source component $C$ we have requirement
$\iota(C)\subseteq o(u)$;
\item[(b)] for each sink component $C$ we have requirement
$o(C)\subseteq\iota(u)$;
\item[(c)] if we have requirements $R\subset R'$ we remove $R'$.
\end{TightDesc}
After (c) one-edge requirements are disjoint with other
requirements, hence multi-edge requirements form a bipartite graph
(in which connections have the form of shared edges).

\section{Minimization algorithms}
\subsection{$1.5$-approximation for \MiED}
\label{15approx}

\subsubsection*{Using DFS}

One can find an equivalent digraph using depth first search starting at
any root node $r$.  Because we operate in a strongly connected graph,
only one root call of the depth first search is required.
This algorithm mimics Tarjan's algorithm for finding strongly connected
components and biconnected components.  As usual for depth first search,
the algorithm forms a spanning tree $T$ in which we have an edge $(u,v)$ if
and only if \Dfs{u} made a call \Dfs{v}.  The invariant is

\begin{center}
(A) if \Dfs{u} made a call \Dfs{v} and \Dfs{v} terminated then 
$T[v]\subset scc_{T\cup B}(u)$.
\end{center}

\begin{figure}[t]
{\small
\begin{tabbing}
0123456789\=012\=012\=012\=012\=\kill
\> \Dfs{u} \\
\> $\{$ \> {\sc Counter} \bec {\sc Counter}+1 \\
\> \> \sci{Number}{u} \bec \sci{LowDone}{u} \bec \sci{LowCanDo}{u} \bec
{\sc Counter} \\
\> \> \ffor each edge $(u,v)$ \ \ \ \  // {\em scan the adjacency list of $u$}\\
\> \>\> \iif \sci{Number}{v} = 0 \\
\> \>\>\> {\sc Insert($T,(u,v)$)} \ \ \ \ \ // {\em $(u,v)$ is a tree edge} \\
\> \>\>\>\Dfs{v} \\
\> \>\>\>\iif \sci{LowDone}{u} $>$ \sci{LowDone}{v} \\
\> \>\>\>\>\sci{LowDone}{u} \bec \sci{LowDone}{v} \\
\> \>\>\>\iif \sci{LowCanDo}{u} $>$ \sci{LowCanDo}{v} \\
\> \>\>\>\>\sci{LowCanDo}{u} \bec \sci{LowCanDo}{v} \\
\> \>\>\>\>\sci{LowEdge}{u} \bec \sci{LowEdge}{v} \\
\> \>\> \eelse \iif \sci{LowCanDo}{u} $>$ \sci{Number}{v} \\
\> \>\>\> \sci{LowCanDo}{u} \bec \sci{Number}{v} \\
\> \>\>\> \sci{LowEdge}{u} \bec $(u,v)$ \\
\> \>  // {\em the final check: do we need another back edge?} \\
\> \> \iif \sci{LowDone}{u} = \sci{Number}{u} {\bf and} $u \not= r$\\
\> \>\> {\sc Insert($B$,\sci{LowEdge}{u})} \ \ \ // \sci{LowEdge}{u} {\em is a
back edge}\\
\> \>\> \sci{LowDone}{u} \bec \sci{LowCanDo}{u} \\
\> $\}$\>  \\ [2ex]
\> $T$ \bec $B$ \bec $\es$ \\
\> \ffor every node $u$ \\
\> \> \sci{Number}{u} \bec 0 \\
\> {\sc Counter} \bec 0 \\
\> \Dfs{r} \\ 
\end{tabbing}
}
\vspace{-6ex}
\caption{\label{dfscode}\dfs\ for finding an equivalent digraph of a strongly connected graph}
\end{figure}

\noindent
(A) implies that $(V,T\cup B)$ is strongly connected when \Dfs{r} terminates.
Moreover, in any depth first search the arguments of calls that already
have started and have not terminated yet form a simple path starting at the
root.  By (A), every node already visited is, in
$(V,T\cup B)$, strongly
connected to an ancestor who has not terminated.
Thus, (A) implies that the strongly connected components of
$(V,T\cup B)$
form a simple path.
This justifies our convention
of using the term {\em back edge} for all non-tree edges.

To prove the invariant, we first observe that when \Dfs{u} terminates
then \sci{LowCanDo}{u}\ is the lowest number of an
end of an edge that starts in $T[u]$.

Application of (A) to each child of $v$ shows that
$T[v]\subset scc_{T\cup B}(v)$
when we perform the final
check of \Dfs{v}.

If the condition of the final check is false, we already have a $B$ edge
from $T[v]$ to an ancestor of $u$, and thus we have a path from $v$ to $u$
in $T\cup B$.
Otherwise, we attempt to insert such an edge.  If \sci{LowCanDo}{v} is
``not good enough'' then there is no path from $T[v]$ to $u$, a contradiction
with the assumption that the graph is strongly connected.

The actual algorithm is based on the above \dfs, {\em but we also need
to alter the set of selected edges in some cases}.

\subsubsection*{Objects, credits, debits}

The initial solution $L$ to the system P3 is divided
into {\em objects}, strongly connected components of $(V,L)$.
L-edges are either inside objects, or between objects.
We allocate L-edges to objects, and give
\Eu{1.5} for each.
In turn, an object has to pay for solution edges that connect it, for a
T-edge
that enters this object and for a B-edge that connects it to an ancestor.
Each solution edge costs \Eu{1}.
Some object have enough money to pay for all $L$-edges inside,
so they become strongly connected, 
and two more edges of the solution, to enter and to
exit.  We call them {\em rich}.  Other objects are {\em poor} and we
have to handle them somehow.

When we discuss a small object $A$,
we call it a {\em path node}, a {\em digon} or a
{\em triangles} when $|A|=1,2,3$ respectively.

\subsubsection*{Allocation of L-edges to objects}

\begin{TightItem}
\item 
$L$-edge inside object $A$: allocate to $A$;
\item 
from object $A$: call the first $L$-edge \Rbr{primary}, and the
rest \Rbr{secondary};
% \marginpar{only two\\ objects take\\
% from secon- dary edges, simpler for $TR_1$}
\begin{TightItem}
\item 
primary $L$-edge \Mat{$A\arr B$}, $|A|=1$: 
\Eu{1.5} to $A$;
\item 
primary $L$-edge \Mat{$A\arr B$}, $|A|>1$: 
\Eu{1} to $A$, and \Eu{0.5} to $B$;
\item 
secondary $L$-edge \Mat{$A\arr B$}:
\Eu{0.5} to $B$ (\Eu{1} to be allocated in the analysis of \MiTRo).
\end{TightItem}
\end{TightItem}

\vspace{-2.5ex}
Later we will formulate Rule \ding{77} to assure a desired property
of primary edges.

\vspace{-2.5ex}
\subsubsection*{When is an object $A$ rich?}

\begin{TightEnum}
\item 
$A$ is the root object, no payment for incoming and returning edges;
\item 
\Mat{$|A|\ge 4$}:
it needs at most $L$-edges inside, plus two edges, and it has
at least \Eu{0.5|A|} for these two edges;
\item 
if $|A|>1$ and an $L$-edge exits $A$:
it needs at most $L$-edges inside, plus two edges, and it has
at least \Eu{(1+0.5|A|)} for these two edges;
\item 
if $|A|=1,3$ and a secondary $L$-edge enters $A$; 
\item 
if $|A|=1,3$ and a primary $L$-edge enters $A$ from some $D$
where $|D|>1$.
\end{TightEnum}

\vspace{-4.5ex}
\subsubsection*{Guiding \dfs}

For a rich object $A$,
% we decide at once to
use $L$-edges inside $A$
in our solution, and we consider it in \dfs as a single node, with combined
adjacency list.  This makes point 
{\bf (1)} below moot.
Otherwise, the preferences are in the order: 
{\bf (1)}
$L$-edges inside the same object;
{\bf (2)}  
primary $L$-edges;
{\bf (3)} 
other edges.

\vspace{-1.5ex}
\subsubsection*{Analyzing the balance of poor objects}

A poor object $A$ has parent object $C$;  \dfs enters
$A$ from $C$ to node $u\in A$.

We say that $A$ shares (the cost of a $B$-edge) if either a $B$-edge to an
ancestor of $C$ is introduced by \dfs within a proper descendant $D$ of $A$
($A$ and $D$ share the cost) or
\dfs from an element of $A$ introduces a $B$-edge to a proper ancestor of $C$
($A$ and $C$ share the cost).  Path nodes and triangles that share 
have needs reduced to \Eu{1.5} and \Eu{4.5} respectively, hence
they achieve balance.

\noindent 
{\bf Case 1: $|A|=1$, $A=\{u\}$}, $A$ does not share.
 
Because we have requirements contained in
\Mat{$\iota(u)$} and in
\Mat{$o(u)$}, there exists $L$-edges that enter and exit $u$.
If an $L$-edge entering $u$ is secondary or exits a
multi-node object, $A$ is rich.
Hence we assume a primary edge $(t,u)$ from object \Mat{$\{t\}$}.

\begin{figure}
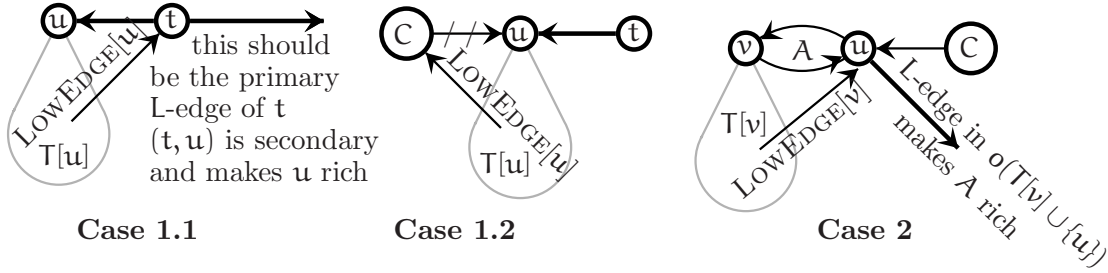

\begin{picture}(430,100)(30,0)
\put(0,30){
\begin{picture}(100,90)(0,0)
\put(30,-10){\epsfig{file=caseonea.eps,width=120pt}}
\put(31,4){\rotatebox{45}{\sci{LowEdge}{u}}}
\put(42,-2){$T[u]$}
\put(100,40){this should}
\put(84,28){be the primary}
\put(84,16){$L$-edge of $t$}
\put(84, 4){$(t,u)$ is secondary}
\put(84,-8){and makes $u$ rich}
\put(46,50){$u$}
\put(90,50){$t$}
\end{picture}
}
\put(60,0){\bf Case 1.1}
\put(120,35){
\begin{picture}(100,70)(0,0)
\put(50,-20){\epsfig{file=caseoneb.eps,width=105pt}}
\put(74,33){\rotatebox{-45}{\sci{LowEdge}{u}}}
\put(89,-10){$T[u]$}
\put(75,40){\small / /}
\put(56,39){$C$}
\put(100,40){$u$}
\put(145,40){$t$}
\end{picture}
}
\put(180,0){\bf Case 1.2}
\put(240,30){
\begin{picture}(100,90)(0,0)
\put(50,-20){\epsfig{file=casetwo.eps,width=115pt}}
\put(63,-15){\rotatebox{40}{\sci{LowEdge}{v}}}
\put(125,34){\rotatebox{-45}{$L$-edge in $o(T[v]\cup\{u\})$}}
\put(125,14){\rotatebox{-45}{makes $A$ rich}}
\put(60,10){$T[v]$}
\put(87,38){$A$}
\put(150,39){$C$}
\put(109,40){$u$}
\put(66.5,40){$v$}
\end{picture}
}
\put(340,0){\bf Case 2}
\end{picture}
\caption{
\label{singles}
Illustrations for the cases of path nodes and digons.
}
\vspace{-2ex}
\end{figure}

\noindent 
{\bf Case 1.1:} \Mat{$C=\{t\}$.} Because $A$ does not share,
no edge to an ancestor of $t$ is present in $B$
when the final check of \Dfs{u} is performed,
and thus $T[u]$ is already strongly connected.  \Mat{\Dfs{u}}
inserts \Mat{\sci{LowEdge}{u}}.  Would this edge go to a proper ancestor
of $t$, $A$ would share (with \Mat{\{t\}}).
Thus 
\Mat{\sci{LowEdge}{u}} goes to \Mat{$\{t\}$} (see \FI{singles}).
Then \Mat{$o(T[u])\subset \iota(u)$},
hence \Mat{$o(T[u]\cup\{t\})\subset o(u)$}, hence there must
be an $L$-edge from $t$ to a node different 
than $u$, a \Rbr{secondary} edge from $t$.

However, we can eliminate this situation by a rule of 
selecting the primary edges.  

\noindent{\bf Rule \ding{77}}: {\em
When \dfs visits a path node object $\{u\}$, it selects a primary edge, an
$L$-edge $(u,v)$ such that \Dfs{v} is the first recursive call of \Dfs{u}, in
such a way that in $V-\{u\}$ a proper ancestor $\{u\}$ is reachable from $v$.}

To see that there exists $(u,v)$ that satisfies Rule \ding{77} 
suppose that $L$-edges $(u,v_i)$, $i=1,\ldots,k$ fail this rule and
$S_i$ is the set of nodes reachable from $v_i$ in $V-\{u\}$.  Then
$R = o(\{u\}\cup S_1\cup\ldots S_k) \subset o(u)$
and $R$ must contain a suitable $L$-edge.

\noindent 
{\bf Case 1.2:} \Mat{$C\not=\{t\}$.}
Initially we pay for $C\arr u$ and \sci{LowEdge}{u}.

This means that \Mat{t} will be visited later.
Because $A$ does not share,
it neither helps connecting $T[u]$ (which is strongly
connected), nor it helps connecting $C$ with its ancestor.

Thus it is OK when
we delete $T$-edge \Mat{$C\arr u$} and we wait until $t$
is visited in the future.  Then \Mat{\Dfs{t}} introduces edge $(t,u)$,
paid by $A$ using the money for the deleted edge, and the cost
of \sci{LowEdge}{u} is shared by $A$ and \Mat{$\{t\}$}.

Note that our actual algorithm differs from \dfs in two ways: \ding{172} \Dfs{t} inserts
to $B$ the primary edge exiting $t$ without waiting for the results of its recursive
calls, and \ding{173} we insert this $L$-edge and delete a $T$ edge.
We will describe similar deviations in the subsequent cases.

\noindent
{\bf Case 2: $|A|=2$, $A=\{u,v\}$.}
\Mat{\Dfs{u}} starts with making the call \Mat{\Dfs{v}}.

We consider what happens when we execute the final check of \Mat{\Dfs{v}}.
If an edge to an ancestor of $C$ is already introduced, $A$
has to pay for $T$-edge \Mat{$C\arr u$}, for edge $(u,v)$ and it
can ``afford'' to pay \Eu{1} for that $B$-edge (more than \Eu{0.5} for
sharing the cost).

If an edge to $u$ is already introduced, $A$ does not share its cost,
while \Mat{$T[v]\cup\{u\}$} is already strongly connected.  At the end
of \Mat{\Dfs{u}}, $A$ can afford to pay for introducing a $B$-edge.

Now we assume that no edge to a proper ancestor of $v$ was introduced before
the final check of \Mat{\Dfs{v}}, and thus $T[v]$ is already strongly
connected.  Thus \Dfs{v} inserts \Mat{\sci{LowEdge}{v}} to $B$.  If this
edge goes to an ancestor of $C$, again, $A$ pays for three edges only.

If \sci{LowEdge}{v} goes to $u$, then $o(T[v])\subset\iota(u)$,
hence $R=o(T[v]\cup\{u\})\subset o(u)$.
Then $R$ contains an $L$-edge hat exits $u$
and does not go to $v$, and this means that $A$ is rich.

\noindent 
{\bf Case 3: $|A|=3$, $A=\{u,v,w\}$.}
While the previous two cases are much simpler then in \cite{V01}, the case
of $|A|=3$ is roughly similar, and we give the details
in Appendix A.

\subsection{Extending the algorithm for \MiED to \MiTRo}
\label{general}

When the set of required edges is not empty, $D\neq\es$, the approach in the
previous section has to be somewhat modified.  When we form ``lower bound''
edge set $L$ we clearly have $D\subset L$, but the algorithm in some cases
fails to include $L$-edges in the solution.
It never happens with $L$-edges in paths and rich objects, so it suffices
to consider poor digons and triangles, and make necessary modification to
our algorithm.

If an L-edge is not ``noticed'' by the algorithm, then it was not considered
in the lower bound used to justify the edges of the solution, so when we
insert this edge, we can also ``notice'' its contribution to the lower bound.

\subsubsection{Digons with $D$-edges}

A {\em problematic digon} consists of a non-$D$-edge $(u,v)$ and a $D$-edge
$(v,u)$.  If a problematic digon can can be adopted as a digon of $L$ and
subsequently it can cause the algorithm for \MiED to ``malfunction'' \IE to
remove its $D$-edge, we say that it is worrisome.  We need to prevent worrisome
digons from being considered as objects by appropriately modified
algorithm for \MiED.

We proceed in two stages.  First, we show how to handle the case of problematic
digons that are exited or entered with $L$-edges (including $D$-edges).
The remaining problematic digons are disjoint.

Because they are disjoint, each of them has to be separately entered and
exited, plus we need to enter the beginning of the $D$-edge of the digon
and exit the end of this edge.  Thus among different ways to enter the
digon (or exit) we value more those that satisfy two requirement rather than
one.  This gives the rise to a maximum weight matching problem.  The details
are in Appendix B.
% If $V-\{u,v\}$ has more than one weak component, then we can replace
% $\{u,v\}$ with a single node $x_{uv}$, and in resulting graph is split into
% two components by $x_{uv}$, so in one of these components we can designate
% the object that contains $x_{uv}$ as the root object, which has surplus
% of \Eu{1}.  This can pay for the edge $(u,v)$.
% 
\subsubsection{Triangles}

Our algorithm for \MiED can be applied to triangles with small modifications.
It is still the case that when a triangle is free we can connect its nodes with
the rest of the solution using 4 edges, but the argument has to be a bit
different in the presence of required edges.

Thus suppose that we obtained a solution for the complement of a triangle
$A=(u,v,w)$ in which edge $(w,u)$ is required.  If we cannot enter $A$ through
node $v$, $v$ has to be entered from inside $A$, hence every solution must have
two edges inside $A$, hence we can collapse $A$ in the preprocessing.
The case when there is no exit of $A$ from node $u$ is symmetric.
Thus we can enter $A$ through $v$, traverse $(v,w,u)$ and exit through $u$.
We say that $A$ is {\em free}, and a free triangle has a surplus of \Eu{0.5}.

One can see that a free triangle without required edges and which is not
collapsed in preprocessing also saves an edge and arrives at a surplus.

We can summarize this section with the following theorem:
\begin{thm}
\label{mitro}
There is a polynomial time algorithms that given an input graph
$(V,E)$ and a set of required edges $D\subset E$ produces a transitive
reduction $H$ such that $D\subseteq H\subseteq E$ and
$|H|\le 1.5k-1$, where $k$ is the size of an optimum solution.
\end{thm}
The reason for $-1$ in the statement is that no edges are added for
the root object in $L$, and this object has at least 2 edges.

\section{$2$-approximation for \MaTRo}

\begin{thm}
\label{matro}
There is a polynomial time algorithms that given an input graph
$(V,E)$ and a set of required edges $D\subset E$ produces a transitive
reduction $H$ such that $D\subseteq H\subseteq E$ and
$|E-H|\ge 0.5k+1$, where $k$ is the size of $|E-H|$ for an optimum solution.
\end{thm}

(In the proof, we add in parenthesis the parts needed to prove $0.5k+1$
bound rather than $0.5k$.)
First, we determine the {\em necessary} edges: $e$ is necessary
if $e\in D$ or $\{e\}=\iota(S)$ for some node set $S$.  (If there are any
cycles of necessary edges, we replace them with single nodes.)

We give a cost of $0$ 
to the necessary edges and a cost of $1$ for the remaining ones.
Remember the primal/dual formulations (in particular (P2) and (D2))
of Section~\ref{p1d1}. 
We set $x_e=1$ if $e$ is a necessary edge and $x_e=0.5$ 
otherwise.  This is a valid solution for the fractional
relaxation of the problem as defined in (P1).  

Now, pick any node $r$.  (Make sure that no necessary edges enter $e$.)
The out-arborescence problem, as defined 
in Section~\ref{p1d1}, is to find a set of edges of minimum cost
that provides a path from $r$ to every other node; edges of cost $0$
can be used in every solution.
An optimum (integral) out-arborescence $T$
can be computed in polynomial time by the
greedy heuristic in~\cite{K72}, this algorithm also provides a set of cuts
that forms a dual solution.

Suppose that $m$ edges of cost 1 are {\em not} included in $T$, then
no solution can delete more than $m$ edges (indeed, more than $m-1$, to
the cuts collected by the greedy algorithm we can add $\iota(r)$).
Let us reduce the cost of edges in $T$ to $0$.  Our fractional solution
is still valid for the in-arborescence, so we can find the in-arborescence
with at most $m/2$ edges that still have cost 1.  Thus we delete at
least $m/2$ edges, while the upper bound is $m$ ($m-1$).

(To assure deletion of at least $\ell/2+1$ edges, where $\ell$ is the
optimum number, we can try in every possible way one initial deletion.
If there optimum number of deletions is $k$, we are left with approximating
among $k-1$, we get an upper bound of at least $k-1$ with $k$ edges left
for possible deletions, so we delete at least $k/2$, plus the initial 1.)

\section{Approximating \MiTR and \MaTR\ for prime $p$}

We will show how to transform our approximation algorithms for \MiTRo
and \MiTRo into approximation algorithms for \MiTR and \MaTR with ratios
1.5 and 2 respectively.  For simplicy. we discuss the case of \MiTR, but
every statement applies to \MaTR as well.

In a nutshell, we can reduce the approximation
in the general case the case of a strongly connected graph, and in a
strongly connected graph we will show that a solution to \MiTRo
can be transformed into a solution to \MiTR by adding a single
edge, and in polynomial time (proportional to $p$) we can find that edge.

In turn, when we run an approximation algorithms within strongly connected
components, we obtain its approximation ratio even if we add one extra
edge (it is actually a property of our algorithms, but in any case one can try
to guess correctly several solution edges and save an additive constant from
the approximation).

Consider an instance $(V,E,\ell,D)$ of \MaTR.
\label{gen-append}
The following proposition says that it suffices
to restrict our attention to strongly connected
components of $(V,E)$\footnote{The authors in~\cite{ADDS} prove their
result only for \MiTR, but the proof applies to \MaTR as well.}.

\begin{Proposition}~{\rm~\cite{ADDS}} 
Let $\rho>1$ be a constant.
If we are given
$\rho$-approximation of \MaTR for each
strongly connected component of $(V,E)$, we can compute
in polynomial time 
a $\rho$-approximation for $(V,E)$.
\end{Proposition}

The following characterization of strongly connected graphs
 appears in~\cite{ADDS}. 

\begin{Lemma}~{\rm~\cite{ADDS}}\label{l1}
Let $(C,E[C],\ell,D)$ is an instance of \MaTR.
Every strongly connected component $C$ of $(V,E)$ 
is one of the following two types:
\begin{description}
\item[(Multiple Parity Component)]
 $|\{a\in{\Bbb Z}_p:~(u,v,a)\in Closure_\ell(E)\}|=p$
for any $u,v\in C$;

\item[(Single Parity Components)] 
 $|\{a\in{\Bbb Z}_p:~(u,v,a)\in Closure_\ell(E)\}|=1$
for any $u,v\in C$.
\end{description}
Moreover, $C$ is a multiple parity component if and only if 
it contains a simple cycle of non-zero parity.
\end{Lemma}

Based on the above lemma, we can use the following approach.  Consider an
instance $(V,E,\ell,D)$ of \MiTR.  For every strongly connected component
$C\subset V$ we consider an induced instance of \MiTRo, $(C,E(C),D\cap C)$.  We
find an approximate solution $A_C$ that contains an out-arborescence $T_C$ with
root $r$.  We label each node $u\in C$ with $\ell(u)=\ell(P_u)$ where $P_u$ is
the unique path in $T_C$ from $r$ to $u$.

Now for every $(u,v)\in E(C)$ we check if $\ell(v)=\ell(u)+\ell(u,v)\mod p$.

If this is true for every $e\in E(C)$ then $C$ is a single parity component.
Otherwise, we pick a single edge $(u,v)$
violating the test and we insert it to $A_C$.
This is sufficient because $A_U$ contains a path $Q$ from $v$ to $r$, and
the cycles $(P_u,(u,v),Q)$ and $(P_v,Q)$ have different parities, hence
one of them is non-zero.

\section{Integrality Gap of the LP Formulation for \MiED\ and \MaED}

\begin{Lemma}\label{gap-lemma}
The primal LP formulation for \MiED and \MaED has an 
integrality gap of at least $4/3$ and $3/2$, respectively.
\end{Lemma}

We use the same construction for \MiED and \MaED.
Our graph will consist of $2n$ nodes.  We first define $2n+2$ nodes
of the form $(i,j)$ where $0\le i < 2$ and $0\le i \le n$.  Later we will
collapse together nodes $(0,0)$ and $(1,0)$ into node $0$, as well as nodes
$(0,n)$ and $(1,n)$ into node $n$.
We have two types of edges: $((i,j),(i,j+1))$ and $((i,j),(i\pm 1,j-1))$.

We get a fractional solution by giving coefficient $0.5$ to every edge.
We need to show that $U\not=V$ and $U\not=\es$ implies $|\iota(U)|\ge 2$.
Suppose $\{0,(0,1),\ldots,(0,n-1),n\}\subset U$; let
$j$ be the least number such that $(1,j)\not\in U$; then $\iota(U)$ contains
$((1,j-1),(1,j))$ and $((0,j+1),(1,j))$.  A symmetric argument holds
if $\{0,(0,1),\ldots,(0,n-1),n\}\subset U$.  In the remaining case,
edge disjoint paths $(0,(i,1),\ldots,(i,n-1),n)$, $i=0,1$, contain edges
from $\iota(U)$.

The cost of this fractional solution is $2n$ (two edges from every of $2n$
nodes, times $0.5$).  We will show that the minimum integer solution
costs $\lceil (8n-4)/3\rceil$.  For this, it suffices to show that no
simple cycle in this graph has the length exceeding $4$ nodes.

If no two consecutive edges in a cycle increase the value of the
second coordinate,
no pair of edges increases the this value, so we have at most two such
values, 
hence at most $4$ nodes.  Alternatively, if
a cycle has two such edges, say the path
$((0,i-1),(0,i),(0,i+1))$, it has to return without using edges
that are incident to $(0,i)$
so it has to use $((0,i+1),(1,i))$
and $((1,i),(0,i-1))$, so it is a cycle of length $4$,
$((0,i-1),(0,i),(0,i+1),(1,i))$.

Every edge of a minimum solution belongs to a simple cycle contained in
that solution; we can start with set $\{0\}$ and extend it using an
edge going from the current set, and a simple cycle that contains that edge;
if we add $k$ edges to the solution we add $k-1$ nodes, and $k\le 4$;
thus the average cost of adding a node must be at least $4/3$ edges.
This completes the proof for \MiED.

When we analyze this example for \MaED, fractional relaxation allows $2n$
deletions while actually we can perform only $\frac{4}{3}n$ of them, so
the ratio is $2\frac{3}{4}=\frac{3}{2}$.
%%\end{prf}

\section{MAX-SNP-hardness Results} 

\begin{thm}
Let $k$-\MiED\ and $k$-\MaED\ be the \MiED\ and \MaED\ problems,
respectively,  restricted
to graphs in which the longest cycle has $k$ edges.
Then, $5$-\MiED and $5$-\MaED are both MAX-SNP-hard.
\end{thm}

Khuller \EA used a reduction from 3-SAT to \MiED.  We are basically
using their reduction, but we restrict its application to sets of clauses
in which each literal occurs exactly twice (and each variable, four times).
It was shown in
\cite{BKS03} that this restriction yields a MAX-SNP hard problem.
Details are in Appendix C.

\section*{Acknowledgments}
The authors thank Samir Khuller for useful discussions.

\newpage

\section*{Appendix A: Algorithm for \MiED, triangle cases}

\noindent 
{\bf Case 3: $|A|=3$, $A=\{u,v,w\}$}

We use the following preprocessing.
If a triple of nodes can be connected with a cycle (triangle) $A$, and it must
contain at least two solution edges,
we contract it to a single node (decreasing
the optimum cost by at least 2), find a solution and then we insert triangle $a$
to the solution (increasing the cost by 3).  Clearly, this
preserves approximation ratio 1.5.  

We assume that $(u,v,w)$ is an oriented cycle, so \Mat{\Dfs{u}} starts with
a call to \Mat{\Dfs{v}}, which starts with a call to \Mat{\Dfs{w}}.

\noindent {\bf Case 3.1:}
$A$ contains an endpoint of a primary $L$-edge.  We can repeat the reasoning of
Case 1, assume that this is edge from a path node $t$ etc.

\noindent {\bf Case 3.2:}
Assume that the solution remains strongly connected when we remove $A$.  We say
that $A$ is {\em free}.  We will show that $A$ can have a surplus of \Eu{0.5}
by traversing $A$ as follows: from $V-1$, two edges inside $A$, back to $V-A$,
so the balance is $4.5-4=$ \Eu{0.5}.

Suppose that all edges $(V-A)\arr A$ enter through the same node, then an
$L$-edge has to enter $A$, and either $A$ is rich or we have Case 3.1.
Similarly, if all edges $A\arr (V-A)$ exit from the same node, $A$ is rich. 

If we can exit from node $w$, use the
path $(V-A)\arr u\arr v\arr w\arr (V-A)$.  Otherwise we can assume
exits from both $u$ and $v$.

If we can enter $A$ at node $v$, use the
path $(V-A)\arr v\arr w\arr u\arr (V-A)$ and 
if we can enter $A$ at node $w$, use $(V-A)\arr w\arr u\arr v\arr (V-A)$. 

\noindent {\bf Case 3.3:}
In the remaining case, no $L$-edge enters or exits $A$, edges enter $A$ at two
nodes, and exit at two nodes, $A$ does not share and $A$ is not free.

One conclusion is that \Dfs{u} must visit other objects besides $A$.  Therefore
$T[u]$ has {\em branches} that extend beyond $A$.  We will consider the
structure of those branches.  A branch is completed when a $B$-edge to a proper
ancestor of its first object is inserted; this merges some scc's, say
$D_1,...,D_k$ into the ancestral scc, so it uses $k+1$ edges outside $D_i$'s.
If such a branch has a surplus we transfer \Eu{0.5} to $A$ which completes its
accounting.  Our goal is to show that such a surplus exists, or we can make
$A$ share, or we can make $A$ free.

\noindent {\bf Case 3.3.1:} $k > 2$.  
Each $D_i$ has a ``local surplus'' of at least \Eu{0.5} after paying for the
incoming edge, and this includes the case of nodes of a digon; a digon
has \Eu{3} and two nodes.  Thus we can pay for the last edge and the branch
still has positive surplus.

\noindent {\bf Case 3.3.2:} $k = 2$ and either $D_1$ or $D_2$ is rich.
The accounting is the same as in Case 3.3.1.  

\noindent {\bf Case 3.3.3:} $D_1$ is a path node, say $t$.  Then there exists
an $L$-edge $(s,t)$ and because $A$ is not rich, $s\not\in A$.  Delete
$A\arr t$ and backtrack from $t$ using $L$-edges until you leave $T[u]$ or you
encounter a cycle object, say $F$.  

If you leave $T[u]$ while backtracking, we can replace $C\arr u$ with the edge
that ``left $T[u]$ backwards''.  Because we removed edge $A\arr t$, this
improves the balance by \Eu{1}.

If you reach a cycle object inside $T[u]$ via a primary edge, we obtain a
branch that goes through rich $F$ and $t$, so we have Case 3.3.1 or 3.3.2.
If we reach $F$ via a secondary edge $F\arr s$, $s$ is ``super rich'' with
\Eu{3} and it can transfer \Eu{0.5} to $A$.

\noindent {\bf Case 3.3.4:} $D_2$ is a path node, say $t$, while
$D_1$ is not.  Because $D_1$ is not rich and not a digon, it is a triangle.
We repeat the reasoning of the previous case, while $D_1$ becomes a free
triangle.

\noindent {\bf Case 3.3.5:} Remaining case: each branch either has
a single scc $D_1$, which is rich (otherwise it is a free triangle),
or two triangle scc's, or two single-node scc's that together form
a digon.

\noindent {\bf Case 3.3.5.1:} Open case: there exists an edge between
$V-T[u]$ and $T[u]-A$.  We will analyze the case of an edge
$V-T[u]\arr T[u]-A$, the other case is symmetric.

It this edge enters some $D_1$, then we insert it and delete two edges,
$C\arr A$ and $A\arr D_1$.

If this edge enters some $D_2$ where $D_1,D_2$ is a pair of triangles,
we insert it, remove $C\arr A$, $A\arr D_1$ and $D_1\arr D_2$ and thus we
make $D_1$ a free triangle, so restoring the connections of
$A\arr D_1\arr D_2$ will save an edge.

If this edge enters some $D_2=\{t_2\}$ where $D_1=\{t_1\}$ and $\{t_1,t_2\}$
is a digon, we insert this edge, remove $A\arr t_1$ and $t_2\arr A$ and
insert an edge $(t_1,s)$ for some $s\not=t_2$.  It is not possible that
exits from $t_1$ are restricted to $t_2$, as we would have
$o(\{t_1,t_2\})\subset o(t_2)$, necessitating an $L$-edge exiting the digon
and the digon would be rich.

If $s\in T[u]$, we can delete $C\arr A$ and save.  Otherwise the progress
is in making the set $T[u]-A$ smaller.

\noindent {\bf Case 3.3.5.2:} Closed case.  The edges between $V-T[u]$
and $T[u]$ must include nodes in $A$.  We can free $A$: there must be at
least two nodes in $A$ that are entered by such edges, otherwise the
edge $C\arr A$ is an $L$-edge, Case 3.1.  Similarly, there must be two nodes
in $A$ from which such edges can exit. So we can repeat the reasoning
of Case 3.2.

\noindent {\bf Remark 1.}
When we discuss subcases of Case 3.3, we assumed that the scc's that
are coalesced when a branch is completed are of one of the ``basic types''.
Actually, they can have a nested structure, following the recursive nature
of \dfs.  If this is the case, we can identify an scc $D_i$ with its root
object.  If the root is rich, then $D_i$ inherits the initial balance if
the root, so it is rich.  If the root is a triangle, $D_i$ inherits its
balance as well, additionally, making $D_i$ free has the same effect as
having free $T[u]$, a subtree rooted by $A$ that is discussed in those cases.
Because we want to gain by making a subtree rooted by a triangle free, we
reduce the problem to that of a smaller subtree with the same property.

The cases of path nodes, 3.3.3 and 3.3.4 are not altered if these path
nodes are roots of larger scc's, and neither is Case 3.3.5.2.

Finally, the open case with digon (3.3.5.1) has to be elaborated when
we have an edge from ``outside'' to the subtree rooted at $D_2=\{t_2\}$.
Our argument was that we can proceed to $t_1$, and if we cannot exit from
$t_1$ to a node different than $t_2$ then the digon $\{t_1,t_2\}$ is rich
as $o(t_1,t_2)\subset o(t_2)$.  This argument does not work if we try
to apply it to $T[t_1]$ and $T[t_2]$ rather than to $t_1$ and $t_2$.

However, if 
as $o(T[t_1]\cup\{t_2\})\subset o(t_2)$ the argument is still valid,
so it remains to address the case when we have an edge from $T[t_1]$
to $T[t_2]-\{t_2\}$.  Then rather then using the edge from outside
of $T[u]$ we change the depth first search as follows: when we reach
$t_1$, we give the edge $(t_1,t_2)$ the last priority. and the resulting
$T[t_1]$ will contain some part of $T[t_2]$ and $t_2$ itself.  
Thus we get a path from $A$ to $A$ that includes at least 3 objects:
$t_1$, $t_2$ and an object that belonged to $T[t_2]$, and this path
delivers a surplus to $A$.

% \noindent {\bf Remark 2.}
% It is somewhat hard to point a problem with Vetta proof.  In fact, we use
% several of his ideas.  Nevertheless, he uses the combinatorial lower bound
% much less aggressively, and as a result he has many cases to consider for
% path nodes and digons.  The first consequence is that it is hard to verify
% claims like "from now on we can assume that there are no digons".  The second
% consequence is that Vetta exercises very little control over edge exchanges
% needed by his proof/algorithm, so his method cannot be extended to \MiTRo.

\section*{Appendix B: Algorithm for \MiTRo, digon cases}

A {\em problematic digon} consists of a non-$D$-edge $(u,v)$ and a $D$-edge
$(v,u)$.  If a problematic digon can can be adopted as a digon of $L$ and
subsequently it can cause the algorithm for \MiED to ``malfunction'' \IE to
remove its $D$-edge, we say that it is worrisome.  We need to prevent worrisome
digons from being considered as objects by appropriately modified
algorithm for \MiED.

% If $V-\{u,v\}$ has more than one weak component, then we can replace
% $\{u,v\}$ with a single node $x_{uv}$, and in resulting graph is split into
% two components by $x_{uv}$, so in one of these components we can designate
% the object that contains $x_{uv}$ as the root object, which has surplus
% of \Eu{1}.  This can pay for the edge $(u,v)$.
% 
\noindent{\bf Case A:}
Suppose that a $D$-edge $e$ exits or enters a problematic digon $A=\{u,v\}$.

\noindent{\bf Case A1:}  $e$ exits $A$.
An $L$-edge that exits a digon makes it rich,
more precisely, it provides this digon with \Eu{1}.

\noindent{\bf Case A2:}  $e$ enters $A$ and it 
is a secondary edge, or it originates in a triangle.

We can change the rule allocating the money of a secondary edge so that the
problematic target gets \Eu{1} (the current rule only allocates
\Eu{0.5} for such an edge to its ending, if the latter is a path node).
Similarly, for a primary edge that exits a triangle it suffices to
give \Eu{0.5} to the triangle, leaving \Eu{1} for the problematic target.

\noindent{\bf Case A3:}  $e$ enters $A$ from 
Thus it suffices to consider cases when problematic digon
$\{u,v\}$ is entered by a primary $e$, from a digon or a path node.

\noindent{\bf Case A3:}  $e$ enters $A$ from a path node, $\{t\}$.
We can show that for an object $B$ that is not a path node,
if there exists a primary edge $u\arr B$, where $u$ is a path node,
we can reduce the expenses of $B$ to \Eu{1}.

\noindent{\bf Case A3.1:}
Suppose that \dfs finds $B$ before $u$.

\noindent{\bf Case A3.1.1:}
Before $u$ is visited, \dfs produces a back edge $b$ for $B$.

\noindent{\bf Case A3.1.1.1:}
The cost of $b$ is shared by $B$ with another object.
Then the expenses of $B$ are \Eu{1.5}, so it needs to obtain \Eu{0.5} from
$\{u\}$.  Later, when \dfs finds $\{u\}$, suppose it is done by traversing
an edge not from the ``primary predecessor'' of $u$, hence 
from some cyclic object $B'$; then $\{u\}$ and $B'$ share the cost of
connecting to an ancestor --- at the time of \Dfs{u},
$scc_{T\cup B}(B)$ is this ancestor.  Inductively, ``compound object''
$B'\cup\{u\}\cup B$ will receive \Eu{0.5} later.

On the other hand, if \dfs finds $\{u\}$ by traversing an edge 
from its ``primary predecessor'' that is also a
path node, say $s$, we apply the same argument ($s$ and $t$ share the cost
and will receive \Eu{0.5} later).  Finally, if $\{t\}$ is entered
via a secondary edge or from a cycle, we collect the promised \Eu{0.5}.

\noindent{\bf Case A3.1.1.2:}
The cost of $b$ is not shared by $B$ with another object.
We have the same reasoning as in Case A3.1.1.1,
but the initial expenses of $B$ are \Eu{2}, while later we can delete the
initial edge used to enter $B$, so the initial cost increases by \Eu{0.5}
and subsequent savings by \Eu{1}.

\noindent{\bf Case A3.2:}
$B$ is visited after $\{u\}$, so \Dfs{B} is the first step of \Dfs{u}.
A back edge from a descendant of $B$ to $\{u\}$ may create $B'$,
and the same applies to a back edge from $B$ to the predecessor of $u$,
while a back edge from $B$ to $u$ implies that there is no other way of
exiting $T[B]$, and, like in Case 1.1, we can conclude that $u\arr B$
is not a primary edge.

\noindent{\bf Case A4:} $e$ enters $A$ from a digon $C$.

\noindent{\bf Case A4.1:} \dfs finds $A$ before $C$.
If $A$ shares the cost
of a back edge with another object, then its expenses are \Eu{1.5}, and when
$C$ is visited, it uses the connecting $L$-edge to its ancestor that contains
$A$, so it pays only for entering; together $C$ and $A$ have
5 edges and have expenses of \Eu{2.5} (not counting the edges that they have
and which they do not delete).  If $A$ makes a back edge without sharing
the cost, then when $C$ is visited we delete the edge used to enter
$A$.  It remains to consider the case when $C$ is visited during
\Dfs(A), and with the ``obligatory'' edge $C\arr A$ this is
closing a circuit that consists of $A,B_1,\ldots,B_k,C$.
This circuit has $k+2$ objects, is closed with $k+1$ edges (not including
the ``obligatory'' edge $C\arr A$), and it needs to be entered and
exited.  The intermediate objects contribute $k-1$ to the needs and
\Eu{1.5k} to the amount of money to pay for these needs, while
$A$ and $C$ contribute \Eu{2.5}, so if $k>0$ we have a balance as
$1.5k+2.5\ge k+4$.  If $k=0$, this means that we enter $C$ with an edge from
$A$.  Note that it is still possible that $A$ will share the
cost of a back edge and we get a balance.  If not, $V-A-C$ is
strongly connected in $T\cup B$, so we can try to improve the solution
by replacing edges $\arr A\arr C$ with $\arr C$.  If not possible,
$C$ can be accessed only from $A$.

Now we have to ask: why is edge $C\arr\{u,v\}$ in $L$?  If this is because of
a requirement that can be satisfied by an edge that leaves $C\cup A$,
then we will make such a replacement, and we now use a ``macro-path''
$\arr A\arr C\arr$, with 2 new edges and 1 edge replaced.  If this is
because of a requirement that must be satisfied by an $C\arr A$ edge, we
have no ``nice'' requirement that requires $A\arr C$ edge; consequently
we can choose an edge $A\arr C$ in such a way that we could delete an
edge inside $C$.  This will not work only if $C$ contains a $D$ edge.
Summarizing, $A\cup C$ must contain $D$ edges inside $A$ and inside $C$, and
also edges $A\arr C$ and $C\arr A$, while we can connect them using 6 edges.
In such a case we collapse $A\cup C$ to a single node, which removes at least
4 edges from an optimum solution, and then we can modify the resulting
solution by adding 6 edges that connect $A\cup C$.

\noindent{\bf Case A4.2:} \dfs finds $C$ before $A$.
Because $C$ is a rich digon,
we treat it as a single node and \Dfs{C} starts by visiting $A$.

If $A$ shares the cost of a back edge with a descendant, $A$ and $C$ together
spend \Eu{2.5} on new edges, and if $A$ shares this cost with $C$, they
spend \Eu{2}.  If $A$ does not share the cost but $C$ is a non-problematic
digon, the combination of edges $C\arr A\arr C$ can remove an edge in $C$.
Lastly, if $C$ is a problematic digon and $A$ cannot share, 
we again have the case that $A\cup C$ must contain at least 4 edges of an
optimum solution and we can collapse $A\cup C$ and connect it using 6 edges.

After this case analysis,
we can conclude that using collapsing of pairs and quadruples we can
eliminate the possibility of worrisome digons that are entered by $D$-edges
or which emanate $D$-edges.  The remaining worrisome digons are disjoint.
Moreover, if $\{u,v\}$ is a problematic digon and we have more than one
requirement contained in $o(u)$, the digon is not worrisome, and the same
is true for $\iota(u)$, $o(v)$ and $\iota(v)$.

Now we are altering the linear program P3 -- which keeps of those
requirements $Rx\ge 1$ of P2 that for some node $u$ satisfy
$R\subseteq o(u)$ or $R\subseteq\iota(u)$, let us call them
The dual D3 is to find a maximum size collection of requirements of P3 such
that no two can be satisfied by a single edge ($i$ such requirements $=$\hspace{-1ex}$\Rightarrow$ we need $i$ edges).

We will form a new dual program, D4.
For a node $w$ that does not belong to a worrisome digon, as before we
introduce requirements contained in $o(w)$ and in $\iota(w)$.
For a worrisome digon with $D$-edge $(v,u)$, we have the one-edge
requirement $\{(v,u)\}$, and thus no other requirements contained in
$o(v)$ and $\iota(u)$.  Also, rather than having requirements
$o(u)$ and $\iota(v)$ we form requirement as follows:
\begin{itemize}
\item
if there is more then one
requirement contained in $o(u,v)=o(\{u,v\})$ we will have these requirements;
\item
if there is only one minimal requirement $R$
contained in $o(u,v)$, we will form a
{\em sibling pair} of requirement: $R$, for simplicity referred to as
as $o(u,v)$, as well as $o(u)$, both with coefficient $0.5$.
\item
if there is more then one
requirement contained in $\iota(u,v))$ we will have these requirements;
\item
if there is only one minimal requirement $R$ contained in $\iota(u,v)$,
we will have a sibling pair of requirements with coefficients $0.5$,
$R$, referred to it as $o(u,v)$, as well as $\iota(v)$.
\end{itemize}

In P4 an edge can satisfy more than two requirements, but the sum of
coefficients satisfied by an edge is at most two. 

Assume that the sum of coefficients of requirements in P4 is $N$
and consider a set of edges $L^\ast$ that satisfies all of them.  For every
requirement we give credit to one of the $L^\ast$-edges that satisfy them.
Say that a net credit of an edge is the sum of credits it received minus one;
then the size of $L^\ast$ is $N$ minus the sum of net credits of edges of
$L^\ast$.

Now, consider a set of edges $M$ of $L^\ast$ with positive net credits; the
lower bound for the size of $L\ast$ is $N$ minus the sum of net credits in
$M$.  We can assign credits in such a way that $M$ can be viewed as a
matching.  

Consider a sibling pair of requirements, $o(u)$ and $o(u,v)$ (the same observation will hold for $\iota(v)$ and $\iota(u,v)$).  The only way to satisfy
the former without satisfying the latter is with edge $(u,v)$; this edge
satisfies only two requirements, both with coefficient $0.5$, so it cannot
belong to $M$.  Thus if an edge in $M$ gets credit for satisfying 
$o(u)$ we can also give it the credit for satisfying $o(u,v)$; in this manner
only one edge can get credit from a sibling pair of requirements.

This defines a bipartite graph: ``nodes'' are requirements with coefficient 1
and sibling pairs, edges are pairs of ``nodes'' in which requirements can
be satisfied simultaneously, edge weights are net credits.
We can find $M$ as a matching in this graph with the maximum weight.

Once we find $M$, we can complete the lower bound by taking account of the
credits not distributed to $M$:

(a) If we have a requirement with coefficient
1 that did not give credit to $M$, we can add an edge that satisfies it,
the same can be done for a sibling pair of requirements that did not
give any credits to $M$ (note that we can satisfy a sibling pair with
a single edge).

(b) In the case of a sibling pair that gave $0.5$ credit to $M$, its digon
retains $0.5$ credit, and we use it to collapse this digon to a single
node: collapsed digon uses two edges, and it has \Eu{1.5} for its
$D$-edge and \Eu{0.5} for its remaining credit.

After applying (a-b), all credits are used up and all requirements of P4
are satisfied, either within the computed lower bound, or using the
``extra credit'' from collapsed $D$-edges.   Moreover, for each new node
$x_{uv}$ (a collapsed worrisome digon) we have edges that satisfy all the 
minimal requirements contained in $o(x){uv}$ and in $\iota(x_{uv})$.

However, we may have the following pathology: in the resulting set of
edges $L$ we may have nodes with no edge exiting (or no edge entering).
For no edge entering, it may happen like that: for a worrisome digon
$\{u,v\}$ with $D$-edge $(v,u)$ we have multiple requirements contained
in $\iota(u,v)$, and we satisfied all of them with edges that enter $u$.
We will try to correct this as follows: if one of these edges exits a
cycle, we replace it with an edge that also satisfies this requirement,
but which enters $v$.  Thus the pathology remains if none of at least
two $L$-edges entering $\{u,v\}$ exits a cycle.  One of these edges can
be on a cycle that contains $u$, but at least one enters $\{u,v\}$ from
a path node (or a worrisome digon that can be considered like a path node).
We can collapse a worrisome digon that exhibits such a pathology, and because
it becomes either a path node with two edges entering from other path nodes,
or a part of a cycle that is entered with an edge from a path node, we
can collect \Eu{0.5} for the use within $\{u,v\}$.

Of course, a similar pathology and its resolution may happen when
we have multiple $L$-edges exiting $\{u,v\}$.

\section*{Appendix C: MAX-SNP-hardness Results} 

\newcommand{\ox}{{\overline{x}}}

\begin{thm}\label{max-snp-hardness}
Let $k$-\MiED\ and $k$-\MaED\ be the \MiED\ and \MaED\ problems,
respectively,  restricted
to graphs in which the longest cycle has $k$ edges.
Then, $5$-\MiED and $5$-\MaED are both MAX-SNP-hard.
\end{thm}

We will use a single approximation reduction that reduces \regsat
to 5-\MiED and 5-\MaED.

In \msat problem the input is a set $S$
of disjunctions of literals, a valid solution is an assignment of truth
values (a mapping from variables to $\{0,1\}$), and the objective function
is the number of clauses in $S$ that are satisfied.
\regsat is \msat restricted to sets of clauses in which every
variable $x$ occurs exactly four times (of course, if it occurs at all),
twice as literal $x$, twice as literal $\ox$.
This problem is MAX-SNP hard even if we impose another constraint, namely
that each clause has exactly three literals~\cite{BKS03}. 

Consider an instance $S$ of \regsat with $n$ variables and $m$ clauses.
We construct a graph with $1+6n+m$ nodes and $14n+m$ edges.
One node is $h$, {\em the hub}.
For each clause $c$ we have node $c$.  For each
variable $x$ we have a gadget $G_x$ with 6 nodes, two switch nodes
labeled $x?$, two nodes that are occurrences of literal $x$ and two
nodes that are occurrences of literal $\ox$.

We have the following edges: $(h,x?)$ for every switch node, $(c,h)$ for every
clause node, $(l,c)$ for every occurrence $l$ of a literal in clause $c$,
while each node gadget is connected with 8 edges as shown in \FI{junk}.

We will show that 
\begin{dingautolist}{172}
\item
if we can satisfy $k$ clauses, then we have a solution
of \MiED with $8n+2m-k$ nodes, which is also a solution of \MaED that
deletes $6n-m+k$ edges;
\item
if we have a solution of \MiED with
$8n+2m-k$ edges, we can show a solution of \regsat that satisfies $k$ clauses.
\end{dingautolist}

To show \ding{172}, we take a truth assignment and form an edge set as follows:
include all edges from $h$ to switch nodes ($2n$ edges) and from clauses
to $h$ ($m$ edges).  For a variable $x$ assigned as true pick set $A_x$
of 6 edges forming
two paths of the form $(x?,\ox,x,c)$, where $c$ is the clause where
literal $x$ occurs, and if $x$ is assigned false, we pick set $A_\ox$
of edges from the paths of the
form $(x?,x,\ox,c)$
($6n$ edges).  At this point, the only nodes that
are not on cycles including $h$ are nodes of unsatisfied clauses, so for
each unsatisfied clause $c$ we pick one of its literal occurrences, $l$
and add edge $(l,c)$ ($m-k$ edges).  

To show \ding{173}, we take a solution $D$ of \MiED.
$D$ must contains all $2n+m$ edges of the form $(h,x?)$ and $(c,h)$.
Let $D_x$ be the subset of $D$ consisting of edges that are incident
to the literals of variable $x$ 
and let $C$ be the set of clause nodes.

Simple inspection of cases show that if $|D_x|=6$ then $D_x=A_x$ or
$D_x=A_\ox$.

If $|D_x|\ge 8$ we replace $D_x$ with $A_x$ and two edges $\ox\arr C$.

If $D_x$ contains $i$ edges to $C$, then $|D_x|\ge 4+i$, because besides
these edges $D_x$ contains 4 edges to the literals of $x$.  If $i=4$ we
are in the case already discussed.  If $i=3$, suppose that a clause
where $\ox$ occurs has no incoming edge in $D_x$; we can replace $D_x$
with $A_x$ plus one edge to a clause in which $\ox$ occurs.  
If a clause where $x$ occurs has no incoming edge in $D_x$, we perform
a symmetric replacement of $D_x$.

In the remaining case $i_x\le 2$ and $|D_x|=7$, and we can perform a replacement
as in the case of $i=3$.

After all these replacements, the size of $A$ did not increase while
each $D_x$ has the form of $A_x$ or $A_\ox$ plus
some edges to $C$.  If $A_x\subset D_x$ we assign $x$ to true, otherwise to
false.   Clearly, if the union of $D_x$'s has $6n+m-k$ edges, at most $m-k$
clauses are not satisfied by this truth assignment (those entered by ``some
other edges to $C$''), so if $|A|=8n+2m-k$, at least $k$ clauses are satisfied.

Berman \EA \cite{BKS03} have a randomized
construction of \regsat instances with
$90n$ variables and $176n$ clauses for which it is NP-hard to tell
if we can leave at most $\eps n$ clauses unsatisfied or at least
$(1-\eps)n$.  The above construction converts it to graphs with
$(14\times 90+176)$ edges in which it is NP-hard to tell if we need
at least $(8\times 90+176+1-\eps)n$ edges or at most
$(8\times 90+176+\eps)n$, which gives a bound on approximability of
\MiED of $1+1/896$, and $1+1/539$ for \MaED.

%%\end{prf}

\begin{figure}[h]
  \begin{center}
    \begin{minipage}[t]{2in}
      \epsfig{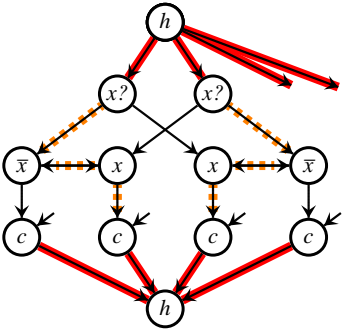}
    \end{minipage}%\hfill
    \begin{minipage}[b]{4in}
       \caption{\label{junk}Illustration of our reduction.
       Marked edges are necessary. 
       Dash-marked edges show set $A_x$ that
       we can interpret it as $x=$true.
       If some $i$ clause nodes are not
       reached (\IE, the corresponding clause is not satisfied) then 
       we need to add $k$ extra edges.  Thus, 
       $k$ unsatisfied clauses correspond to $8n+m+k$ edges being used
       ($6n-k$ deleted) and 
       $k$ satisfied clauses correspond to $8n+2m-k$ edges being used
       ($6n+m-k$ deleted).}
    \end{minipage}
  \end{center}
\end{figure}

% end of the exact algorithm

\hfill$\Box$

\end{document}